\documentclass[12pt]{article}
\usepackage{mathtext}
\usepackage{graphicx}
\usepackage[rightcaption]{sidecap}
\usepackage{amsfonts}
\usepackage{amssymb}
\usepackage{amsmath}

\usepackage{epstopdf}
\usepackage{float}
\usepackage{braket}
\usepackage{indentfirst}

\usepackage[a4paper,margin=2.5cm]{geometry}
\begin{document}
\title{Distributed synthesis of chains with one-way biphotonic control}

\author{Y.I.Ozhigov\thanks{Moscow State University of Lomonosov, VMK Faculty, Institute of Physics and technology RAS (FTIAN), e-mail: ozhigov@cs.msu.su}
}

\maketitle
\begin{abstract}
An example of a one-way distributed computation is given in which the use of entangled states of two photons to synchronize processes gives a benefit. The process of assembling polymer chains at two remote points is considered; the quality of the assembly is determined when they are superimposed on each other. The effect of using entangled states is almost 14 percents.
\end{abstract}

\begin{figure}
\begin{center}
\includegraphics[height=0.5\textwidth]{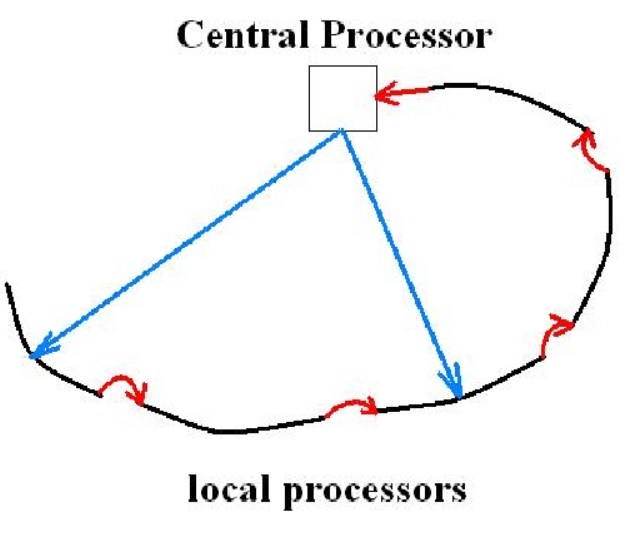}
\caption{One-way control scheme }
\label{fig:nps}
\end{center}
\end{figure}

\section{Introduction. One way control}

The idea of ​​applying quantum mechanics to computations began with the Feynman's project of the quantum computer \cite{Fe}, and then got a powerful development in the theory of quantum computing, including quantum speedup (\cite{Sh}, \cite {Gr}), quantum modeling (\cite{Za}, quantum eror correction (\cite{Sh2}). At the same time, it turned out that the quantum speedup encompassed only algorithms of the search type  (\cite{Oz}), and the serious problem of decoherence (\cite{Dec}) lies on the way of practical realization of quantum computations in general.

In this connection, limited models of quantum computations play the significant role, for example branching quantum programs (\cite{A}) or programs for modeling biochemistry (\cite {Do}.) Here, the advantage of quantum methods can not consist in obtaining global speedup, but in the final effect of using separate elements of essentially quantum nature to obtain the ultimate win.

We will show how this goal can be achieved by quantum nonlocality consisting in violation of Bell inequality (\cite{Bell},\cite{Bell2}). We consider a model of computation with one-way control, where the system is divided into the central processor (CPU) and remote peripheral devices that can directly receive the signal from CPU. Feedback from peripheral devices does not reach the CPU immediately, but only by its sequential transfer through a chain of peripheral devices, which interacts with each other locally as shown in the Figure 1.  The entangled states of photons used to control give a gain in efficiency in about $1.138$ over classical control for the problem of chain synthesis in two remote peripherals.

The central processor sends a signal to local processors, each of which is responsible for the corresponding subsystem in the whole system. For example, the CPU decides to synthesize some polymer $ A $, which has a specific activity in some subsystem, and, at the same time, activates the synthesis of another polymer $ B $ that suppresses (or intensifies) this activity in another subsystem. The CPU sends the corresponding signals to both subsystems, and then switches to another task, for example, to a simultaneous synthesis of another pair of polymers $ A '$ and $ B' $.

What would happen if the subsystems themselves start sending signals to each other? Suppose we have $ m $ subsystems, each of which is controlled by its own processor. For correct addressing of all signals between all possible pairs (there are about $ m ^ 2 $ ) we will have to load the CPU with this work. The CPU will thus be forced to wait for $ cD $, where $ D $ is the distance between local processors, $ c $ is the speed of light, before switching to another job.
If $ m $ is large enough (in real biosystems this number is very large), such a scheme of addressing through the CPU will lead to a fatal delay that makes the whole model uncompetitive. Therefore, the idea of ​​direct signal exchange between local processors is bad.

We, therefore, come to the need for one-way control, in which the CPU sends signals to all 
peripheral processors simultaneously, without waiting for a response from them. The CPU also receives feedback from the local processors not directly, but through a chain of intermediaries, as in a cellular automaton. This form of information processing can be effective in living systems, since here we offload the CPU from carrying out routine work on the actual organization of metabolism.

\section{Quantum biphotonic signals}

Here, there are situations when a CPU using biphotons - entangled photon states - will have an advantage over a fully classical CPU. To demonstrate this possibility, consider the following abstract problem. Let us synthesize two molecules of polymers, which chemical structure looks as $C_1=(c_1^1,c_2^1,...,c_M^1),\ C_2=(c_1^2,c_2^2,...,c_M^2)$, 
and they consist of mono-blocks of two types: $ a $ or $ b $: $c_i^j\in\{ a,b\}$ (see the Figure 2). 

The quality of the joint assembly of the two polymers is checked when they are superimposed on each other - the first $C_1$ on the second $C_2$, and the quality criterion is the degree of their gluing.
 Each mono-block has an outer surface (convex) and internal (concave) where the last one is supplied by the special ball in its center. Two monoblocks in the fixed positions can glue to each other in the following two cases: 1) their surfaces or the half of their surfaces can be strictly combined by the vertical shift, or their central balls are at the same point - as shown in the Figure 2. 

The physical structure of the polymer, which gluing depends on, is not determined only by the sequence of monoblocks; it also depends on the options of their location relative to each other. The monoblocks in the chain are connected by flexible links so that any link can be either compressed by $dx$ that is the quarter of length of the monoblock or stretched by the same value; we say that the monoblock is shifted back or forward relatively to the equilibrium position of the link. During the synthesis, links are established according to these requirements, and then fixed. When two chains are overlayed, the position of the monoblocks one on another is thus fixed. In particular, when two monoblocks with the same number are shifted in the opposite directions (one is shifted back and another forward) the resulting shift will be by the half of the monoblock length. 

Two chains can be then superimposed by these rules and we can count the number of pairs of good gluing monoblocks in it. 

Synthesis of the polymer chain goes by successive attachments to the existing chain of a new monoblock - one that appeared at the growth point first (monoblocs are taken from the surrounding polymer medium, and are there in a chaotic motion). In this case, it is possible to move the monoblock either back or forward by a distance $ dx $. We denote the shift forward through $ + $, the shift back through $ - $. Any $j$-th segment of superimposed chains is thus a quad $ c_j^1c_j^2s_j^1s_j^2$  where the last two elements are shifts $ s_j^{1,2} \in \{ +, - \} $.

It follows from our rules (see Figure 2) that the good gluing corresponds to the pairs of superimposed monoblocks of the form: $aa++(--),\ ab++(--),bb++(--),ba+-(-+)$ whereas pairs of the other form:
$aa+-(-+),ab+-(-+),bb+-(-+),ab++(--)$ give bad gluing. We see the non symmetric behavior of $a$ and $b$ monoblocks: the pairs $ab$ and $ba$ glues differently after the same shifts. This asymmetry looks like asymmetry of Bell inequality, that will give us the advantage of gluing quality of two chains assembled with bi-photons over the classically assembled chains. 

Suppose that the growth of the polymer $ C_1 $ occurs at one point, and the growth of the polymer $ C_2 $ in the other, and these points are separated from each other by a large distance (for example, they are in different countries). The problem is to provide this double synthesis so that the number of critical (non-sticky) sites in a pair of polymers is minimal. In other words, we want to ensure the maximum degree of gluing of two chains synthesized far apart.
\begin{figure}
\begin{center}
\includegraphics[scale=0.7]{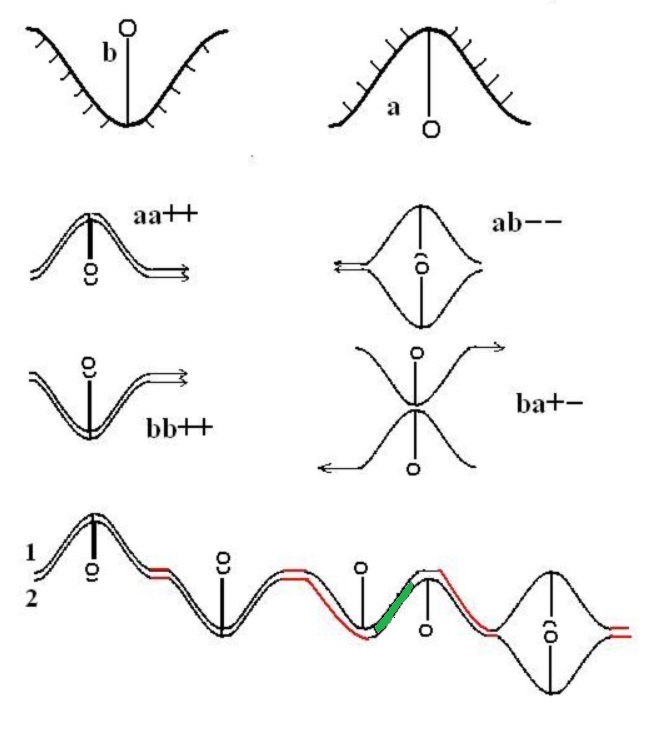}
 \caption{Overlay of two polymers. Arrows indicate the direction of stretching the connections (red color) between mono blocks in the polymer syntethis. Impositions $aa++(--),ab++(--), bb++(--), ba+-(-+)$ give good gluing, the other give bad gluing. At the bottom all pairs of monoblocks glue well. }
\end{center}
\end{figure}

Such a problem can arise when modeling the synthesis of a gene and an antigen in different living cells. We could organize a constant information exchange between points of growth. But if the distance between them is large, such an exchange would essentially slow down the synthesis process, since at each point it would be necessary to wait for the signal to arrive from another point. Here we are still distracted from the problem of organizing such waiting period, which in the real system presents a separate difficult task.

But it is possible to organize one-way control over the process of double synthesis with the help of biphotonic pairs, which gives an increase in the synthesis efficiency by an appreciable amount in comparison with the classical one-way control.

So, in order to minimize the total number of critical overlaps, we want to use CPU signals in the form of EPR states of the type $ | \Psi \rangle = \frac{1}{\sqrt 2} (| 00 \rangle + | 11 \rangle) $, and must count the number obtained in such a control of critical pairs, to show the advantage of control through entangled biphotonic pairs. To do this, we first recall the Bell inequality, which is valid for classical correlations.

 Let we be given four random variables $b_2,b_1,a_1,a_2$, taking values from the set $\{ 1,-1\}$. Let us also suppose that the random values $a_1,\ b_1$ are determined on the set of elementary events separated from the domain of $a_1,\ b_2$ (this is called a local realism, see \cite{Bell2}).  For their mean values the following Bell inequality will then takes place: 
\begin{equation}
E(a_1b_2+b_1b_2+a_1a_2-b_1a_2)\leq 2, 
\label{Bell}
\end{equation}
which in view of local realism follows from the representation   
\begin{equation}
\label{real}
a_1b_2+b_1b_2+a_1a_2-b_1a_2=a_1(b_2+a_2)+b_1(b_2-a_2), 
\end{equation}
because one of expressions in brackets equals zero and the other equals 2. 
  
We accept the following agreements. The subscript denotes the assembly point number: 1 or 2. The letter $ a $ or $ b $ denotes the monoblock type, the sign corresponding to the shift direction is: $ + $ - forward shift, $ - $ - back shift of the current monoblock when it is attached to the chain. The result of joining monoblocks at both points of the assembly is determined if for each lower index 1 and 2 we have, firstly, the letter $ a $ or $ b $, and secondly, the shift sign $ + $ or $ - $. The letter $ a $ or $ b $ always determines the type of monoblock, which at the time of assembly is closest to its point.

The CPU can thus control the assembly only by selecting the shift sign: $ + $ or $ - $ at both points of the assembly. In this case, the CPU selects these symbols simultaneously, so that no waiting for a signal from one assembly point to the other can slow down the process. Note that if allow delay on time, then the quality of gluing can be made ideal. 

For the case of the classical correlation between the choice of sign we have the Bell inequality. For each step of the process, we introduce the criticality index $ Cr = + 1 $, if the imposition of the monoblocks obtained at this step turned out to be uncritical, that is they will glue together when superimposed, and $ Cr = -1 $ otherwise. We are interested in the total number of non-critical impositions in the whole pair of synthesized polymers: $ NonCr $. Our goal is to make this number the maximum.

 For one pair of monoblocks we have $NonCr=\frac{1}{2}(1+Cr)$. Since all combinations $aa, ab, ba, bb$ for both points of the assembly have equal probabilities $1/4$, for the mean value $E(Cr)$ of the criticality index we have 
\begin{equation}
E(Cr)=\frac{1}{4}(a_1b_2+b_1b_2+a_1a_2-b_1a_2)
\label{cr}
\end{equation}
where the letter $a$ or $b$ with the index denotes the random variable, corresponding to the choice of monoblock type with the value $\pm 1$ depending on what sign of shift is chosen for it.
In the case of classical control in view of Bell inequality for $E(Cr)$ of the form \eqref{cr} the mean value of critical impositions satisfies the inequality $E(NonCr)\leq \frac{1}{2}(1+\frac{2}{4})=\frac{3}{4}=0.75$. 

In the case of biphotonic control, the situation will be different. Here we cannot consider $a_{1}$ or $b_{1}$ as random variables determined on the set of events separated from the domain of $a_2,\ b_2$, e.g., the evaluation \eqref{Bell} does not now follow from the evident expression \eqref{real}: we must write $a_1b_2+b_1b'_2+a_1a_2-b_1a'_2$ instead of the right side of trivial equality \eqref{real} that would make it wrong and we cannot so obtain Bell inequality \eqref{Bell}. 

For biphotonic control our random valiables are determined on the same set of elementary events, which means that here we will not have Bell's inequality and must find the probabilities directly using the Born rule.

Suppose that at each assembly point there is an imaginary photodetector that can be instantaneously oriented in accordance with the observables associated with $ a $ and $ b $. For the first and second assembly points, these observables will have the form

\begin{equation}
\label{ob}
\begin{array}{lll}
&a_1=\sigma_x,\ &b_1=\sigma_z,\\
&a_2=\frac{1}{\sqrt 2}(\sigma_z-\sigma_x),\ &b_2=-\frac{1}{\sqrt 2}(\sigma_x+\sigma_z)
\end{array}
\end{equation}
correspondingly. Here we do not consider the interesting question of the practical implementation of such observables.

We agree that the type of the current monoblock during assembly determines the corresponding position of the detector in each of the two points, and the sign of the monoblock shift is the value of the corresponding observable. Again, since all combinations of types of monoblocks $ aa, ab, ba, bb $ are equally probable, we can use the formula \ref{cr} for the average value of the criticality index.

We have now: $E=E(a_1b_2+b_1b_2+a_1a_2-b_1a_2)=E(a_1b_2)+E(b_1b_2)+E(a_1a_2)-E(b_1a_2)$. Using the definition of observables \eqref{ob} and applying the rule for finding the mean value $\langle A\rangle_{\psi}=tr(A\rho_{\psi})$ for all observables $A$ taken from \eqref{ob}, we find $E= 2 \sqrt{2} $ and for the average value of the total number of non-critical overlaps, we get the value $ E (NonCr) = 
\frac{1}{2} (1+ \frac{2 \sqrt{2}}{4}) \approx 0.85 $. Thus, the use of EPR pairs of photons to control the assembly gives a tangible gain in the quality of the assembly - slightly more than $1.138$ in this formulation of the problem.
 
\section{Conclusion}

We have shown how the one-way control via entangled biphotons can give the more effective assembly of two polymers in the different points in comparison with classical one-way control. This effect in the abstract form can be demonstrated with the existing computers with the source of biphotons and detectors. 

The effect found seems to be insignificant - about $1.138$, but if we repeat the assembly repeatedly, alternating with other processes, the effect could become significant. The considered situation is probably typical and can occur in a variety of non-equilibrium processes in a living cell or in bacterial populations where the entangled photonic states could play the valuable role. 

\section{Acknowledgements}

This work was partially supported by Russian Foundation for Basic Research
(RFBR) grant a-18-01-00695.

\end{document}